\documentclass[prc,showpacs,floatfix,]{revtex4}
\usepackage{graphicx,subfigure}
\usepackage{comment}
\usepackage{amsmath, amsthm, amssymb}
\usepackage{bm}

\def\nuc#1#2{\relax\ifmmode{}^{#1}{\protect{#2}}\else${}^{#1}$#2\fi}

\newcommand {\la} {\langle}
\newcommand {\ra} {\rangle}
\newcommand {\beq} {\begin{eqnarray}}
\newcommand {\eeq} {\end{eqnarray}}
\newcommand {\eeqn} [1] {\label{#1} \end{eqnarray}}%
\newcommand {\eol} {\nonumber \\}
\newcommand {\ve} [1] {\mbox{\boldmath $#1$}}

\begin{document}
\graphicspath{{figures/}}

\title{Translation invariance and antisymmetry in the theory of the nucleon optical model.\footnote{Phys.Rev.C Accepted version.}}
 \author{R.C. Johnson}
 \affiliation{
  Department of Physics,
 Faculty of Engineering and Physical Sciences,
 University of Surrey,
 Guildford, Surrey GU2 7XH, United Kingdom}

\date{\today}

\begin{abstract}
The first step in any formalism that aims to connect a many-nucleon theory of nucleon-nucleus
scattering and the concept of an optical model potential in the sense pioneered by Feshbach is to
explain what is meant by the optical model wavefunction. By definition this is a function of a single
space coordinate plus a set of single nucleon internal variables. This article gives a critique of the definition
 as it is frequently expressed in 2nd Quantisation language and suggest a new definition which is more consistent with the
requirements of antisymmetry and translational invariance. A modification of the time-dependent Green's function formalism is suggested.
\end{abstract}
\maketitle

\section{Introduction.}\label{introduction}
Theories devoted to calculating an optical potential that describes nucleon scattering from an $A$-nucleon target in terms of  fundamental 2- and 3-body inter-nucleon interactions are of considerable current interest. For scattering from light nuclei fully antisymmetrised and translationally invariant resonating-group \emph {ab initio} calculations have been successfully developed, see \cite{RQN14},\cite{NQ12} and references therein. For heavier targets a recent report by Idini, et al,\cite{IdiniBarbieriNavratil16} references several approaches.  They also report their own calculations of nucleon optical potentials based on a self-consistent Green's function formalism (SCGF) which it is claimed is well  suited to calculating optical potentials for medium mass targets. 

Work on the fundamental definition of the optical model potential goes back to the pioneering work of Feshbach\cite{Fes58}. This work and subsequent work by  Bell and Squires \cite{BellSquires} and Capuzzi and Mahaux\cite{CapuzziMahaux95}, as well as  the more recent work reported in \cite{Rotureau16} and \cite{IdiniBarbieriNavratil16}, all seek a function of one spatial coordinate and a set of single nucleon spin and iso-spin coordinates that can be identified as an optical model wave function. This is usually defined as the projection of the exact antisymmetrised many-body nucleon-$A$ scattering wave function, $\mid \Psi_{E^+}\ra$, onto the exact antisymmetrised target ground state, $\mid  \Psi_0 \ra $ .  It is standard to interpret this definition in 2nd Quantisation notation through the formula
 \beq
\xi_{E^+}(\ve{r})= \la \Psi_0 \mid \psi(\ve{r}) \mid \Psi_{E^+}\ra.\label{chirdef}\eeq
      
For a translationally invariant many-nucleon Hamiltonian the $(A+1)$-nucleon scattering state $\mid \Psi_{E^+}\ra$ can be assumed to have a definite total momentum and in the c.m. system this momentum will be zero. Similarly, it can be assumed that the $A$-nucleon target ground $\mid  \Psi_0 \ra $  also has total momentum zero. Under these assumptions it will be shown that  $\xi_{E^+}(\ve{r})$ as defined in eq.(\ref{chirdef}) is independent of the space coordinate element of $\ve{r}$ and hence cannot possibly be an acceptable definition of the optical model wave function. It is the purpose of this paper to justify this statement and present a modified definition that corresponds more faithfully to the properties expected for the microscopic optical model. No new numerical calculations using the new definition  are presented.
\subsection{The case of single determinant states.}
For medium and heavy targets all approaches have in common the use of a single particle mean-field basis in which nucleon antisymmetry is fully taken into account using second-quantisation techniques but introduces a violation of translation invariance very early in the development. To see how problems arise with this approach  the mean-field limit is considered. In this limit the target ground state  $\mid \Psi_0\ra$ is a single determinant of $A$ bound single particle states in some potential well and  $\mid \Psi_{E^+}\ra$ is a determinant of the same $A$ states but with the extra nucleon in a continuum scattering state, $u_{\ve{k}}(\ve{r})$ in the same potential well.

In this special case the formula (\ref{chirdef}), $\la \Psi_{0}\mid \psi(\ve{r}) \mid \Psi_{E^+}\ra$,  gives precisely the  scattering state $u_{\ve{k}}(\ve{r})$, so identifying the optical potential with the mean field, as expected. This result gives a false sense of security for the validity of formula (\ref{chirdef}) because it fails to recognise the properties of this expression when exact, rather than model many-nucleon wave functions are used. 

 Formally, this  misleading result arises because determinants are not eigenstates of definite total momentum and hence the result outlined in the previous Section and  proved in Section \ref{why} does not apply. If one tried to improve the determinants for their violation of translation invariance by, \emph{e.g.}, projecting them on to their zero total momentum components and then using the formula  (\ref{chirdef}) the result would be a function independent of $\ve{r}$ and certainly not acceptable.
\section{Why the definition of the optical model wave function through eq.(\ref{chirdef}) must be rejected.}\label{why}   
(i) The quantity  $\xi_{E^+}(\ve{r})$ from eq.(\ref{chirdef}) can be evaluated using the wave functions defining $\Psi_0$ and $\Psi_{E^+}$ in configuration space:  
\beq
\xi_{E^+}(\ve{r})&=& \la \Psi_0 \mid \psi(\ve{r}) \mid \Psi_{E^+}\ra\eol
&=&\sqrt{(A+1)}\int d\ve{r}_1\dots d\ve{r}_A\Psi^*_0(\ve{r}_1,\dots\,\ve{r}_{A})\eol &\times &\Psi_{E^+}(\ve{r}_1,\dots\,\ve{r}_{A},\ve{r}_{(A+1)}=\ve{r})\eol
&=&\sqrt{(A+1)}\int d\ve{r}_1\dots d\ve{r}_A d\ve{r}_{(A+1)}\Psi^*_0(\ve{r}_1,\dots\,\ve{r}_{A})\eol &\times &\delta(\ve{r}_{(A+1)}-\ve{r})\Psi_{E^+}(\ve{r}_1,\dots\,\ve{r}_{A},\ve{r}_{(A+1)}).\label{chirdef2}\eeq
 This integral is independent of $\ve{r}$. This can be seen by first translating the first $A$ variables of integration by $\ve{r}$ to $\ve{r}'_i= \ve{r}_i-\ve{r},\,i=1,\dots,A$, and using the fact that $\Psi_0$ and $\Psi_{E^+}$ are zero-momentum states so that
\beq \Psi_0(\ve{r}_1+\ve{r},\dots\,\ve{r}_{A}+\ve{r})&=&\Psi_0(\ve{r}_1,\dots\,\ve{r}_{A}),\eol
\Psi_{E^+}(\ve{r}'_1+\ve{r},\dots\,\ve{r}'_{A}+\ve{r},\ve{r}_{(A+1)})&=&\Psi_{E^+}(\ve{r}_1,\dots\,\ve{r}_{A},\ve{r}_{(A+1)}-\ve{r}).\label{zeromom}\eeq
Hence
 \beq
\xi_{E^+}(\ve{r})
&=&\sqrt{(A+1)}\int d\ve{r}'_1\dots d\ve{r}'_A d\ve{r}_{(A+1)}\Psi^*_0(\ve{r}'_1+\ve{r},\dots\,\ve{r}'_{A}+\ve{r})\eol &\times &\delta(\ve{r}_{(A+1)}-\ve{r})\Psi_{E^+}(\ve{r}'_1+\ve{r},\dots\,\ve{r}'_{A}+\ve{r},\ve{r}_{(A+1)})\eol
&=&\sqrt{(A+1)}\int d\ve{r}'_1\dots d\ve{r}'_A d\ve{r}_{(A+1)}\Psi^*_0(\ve{r}'_1,\dots\,\ve{r}'_{A})\eol &\times &\delta(\ve{r}_{(A+1)}-\ve{r})\Psi_{E^+}(\ve{r}'_1,\dots\,\ve{r}'_{A},\ve{r}_{(A+1)}-\ve{r})\eol
&=&\sqrt{(A+1)}\int d\ve{r}'_1\dots d\ve{r}'_A \Psi^*_0(\ve{r}'_1,\dots\,\ve{r}'_{A})\eol &\times &\Psi_{E^+}(\ve{r}'_1,\dots\,\ve{r}'_{A},\ve{r}_{(A+1)}=0),\label{chirdef2}\eeq
which is independent of $\ve{r}$.

(ii) The same $\ve{r}$-independent result is obtained when the right-hand-side of eq.(\ref{chirdef}) is evaluated using Fock-space formalism techniques. For simplicity  explicit reference to spin and iso-spin coordinates is omitted. The notation $\mid\dots \ra\ra$ is used for vectors in Fock space. 

The states\ $\mid \Psi_{E^+}\ra\ra$ and $\mid \Psi_{0}\ra\ra$
are both eigenstates of the total momentum operator $\hat{\ve{P}}$ with eigenvalue zero, where
 \beq
\hat{\ve{P}}=\int d\ve{k}\, \ve{k}\, a^\dagger_{\ve{k}} a_{\ve{k}},\label{P}\eeq
and $ a^\dagger_{\ve{k}}( a_{\ve{k}})$ creates (destroys) a nucleon in a single particle state with momentum $\ve{k}$.

In terms of state vectors and operators in Fock space 
\beq
\hat{\ve{P}}\mid \Psi_0 \rangle \rangle&=& 0,\eol
\hat{\ve{P}}\mid \Psi_{E^+}\rangle \rangle&=&0
.\label{PPsi0E}\eeq
It follows from  from eq.(\ref{P}) that
\beq [\hat{\ve{P}},\psi(\ve{r})]_-=\imath(\nabla_{\ve{r}}\psi(\ve{r})),\label{Parcomm}\eeq
and hence
\beq \imath(\nabla_{\ve{r}}\xi_{E^+}(\ve{r}))&=&\langle \langle \Psi_0 \mid [\hat{\ve{P}},\psi(\ve{r})]_- \mid \Psi_{E^+}\rangle \rangle\eol
    &=&\langle \langle \Psi_0 \mid (\hat{\ve{P}}\psi(\ve{r})-\psi(\ve{r})\hat{\ve{P}}) \mid \Psi_{E^+}\rangle \rangle\eol &=&  0.\label{Pchi}\eeq
 Hence, $\xi_{E^+}(\ve{r})$ is independent of $\ve{r}$ and can not be the required optical model wave function.
 
 More generally, if $\mid \Psi_{1}\ra\ra$ and $\mid \Psi_{2}\ra\ra$ have momenta $\ve{K}_1$ and $\ve{K}_2$, respectively, then
 \beq \la \la \Psi_1 \mid \psi(\ve{r})\mid \Psi_{2}\ra\ra
 &=&\exp(\imath(\ve{K}_2-\ve{K}_1).\ve{r})\langle \langle \Psi_1\mid \psi(0)\mid \Psi_2\ra\rangle.\label{RotOverlap}\eeq

One comes to the same conclusion  working  in momentum space:
\beq
\xi_{E^+}(\ve{r})&&= \la \Psi_{0}\mid \psi(\ve{r}) \mid \Psi_{E^+}\ra\eol
&&=\int d\ve{k}  \frac{\exp(\imath \ve{k}.\ve{r})}{(2\pi)^{3/2}}\la \Psi_{0} \mid a_{\ve{k}} \mid \Psi_{E^+}\ra\eol
&&=\int d\ve{k}  \frac{\exp(\imath \ve{k}.\ve{r})}{(2\pi)^{3/2}}\la \Psi_{0} \mid a_{\ve{k}=0} \mid \Psi_{E^+}\ra\delta(\ve{k})\eol
&&= \frac{1}{(2\pi)^{3/2}}\la \Psi_{0} \mid a_{\ve{k}=0} \mid \Psi_{E^+}\ra.\label{chirdefmom}\eeq
Again it seen that $\xi_{E^+}(\ve{r})$ as defined by eq.(\ref{chirdef}) is independent of $\ve{r}$.
 The result (\ref{RotOverlap}) follows in a similar fashion. 

     \section{Definition of the optical model wave function.}\label{optwfdef1}
The standard way of defining the type of $A$-nucleon, $(A+1)$-nucleon overlap of which the optical model wave function is an example is to use  a set of $A-1$ translationally invariant variables $\ve{\chi}_1,\dots\,\ve{\chi}_{(A-1)}$ (\emph{e.g.,}  Jacobi coordinates) that together with the $A$-nucleon c.m., $\ve{R}_A$, form a set of variables equivalent to the vectors $\ve{r}_1, \dots \ve{r}_A$ describing the position of the nucleons relative to an arbitrary origin and have a transformation Jacobian equal to  $+1$. The variables $\ve{\chi}_1,\dots\,\ve{\chi}_{(A-1)},\ve{\chi}_A$, together with the $A+1$-nucleon c.m. $\ve{R}_{(A+1)}$, perform the same role for $A+1$ nucleons,  where $\ve{\chi}_A$ is defined by
\beq\ve{\chi}_A= \ve{r}_{(A+1)}-\ve{R}_A,\,\,\,\,\,\ve{R}_A=(\sum_{i=1}^{i=A}\ve{r}_i)/A.\label{chiAdef} \eeq

In terms of these coordinates  the relevant overlap is
\beq \xi_{E^+}(\ve{r})=\sqrt{(A+1)}&&\!\!\!\!\!\int d\ve{\chi}_{1}\, \dots d\ve{\chi}_{(A-1)} \Psi^*_0(\ve{\chi}_1,\dots\,\ve{\chi}_{(A-1)})\eol&&\times \Psi_{E^+}(\ve{\chi}_1,\dots\,\ve{\chi}_{(A-1)},\ve{\chi}_{A}=\ve{r}).\eol && \label{optwfdef}\eeq
The zero momentum functions $\Psi_{E^+}$ and $  \Psi_0  $  are independent of, respectively, $\ve{R}_{(A+1)}$ and $\ve{R}_A$. 

The function defined in eq.(\ref{optwfdef}) is certainly not generally independent of $\ve{r}$. From the definition of $\ve{\chi}_A$ it is clear the the meaning of $\ve{r}$ is the vector distance of the extra nucleon from the c.m. of the target.

 Similar overlap functions are familiar from the study of the overlaps between bound state of the $A$- and $(A+1)$-nucleon systems in the theory of $(d,p)$ and $(p,d)$ reactions. They have been shown to satisfy an inhomogeneous differential equation (the "source equation")  in which the kinetic energy operator appears with the correct reduced mass and whose solutions have been extensively studied, see \cite{Tim11}, \cite{Tim14}. 
 
 The Jacobi coordinates do not lend themselves well to treating antisymmetrised functions. The 2nd Quantisation formalism is more attractive from this point of view. 
In the next Section a new formula is derived for the optical model wave function, and in fact for a general $A$-nucleon, $(A+1)$-nucleon overlap, in terms of operators and states in Fock space by working directly from eq.(\ref{optwfdef}).

\section{The optical model wave function as a matrix element in Fock space.}\label{FockSpaceformulae}
It is straightforward to convert the expression (\ref{optwfdef}) to a relation between state vectors and creation and destruction operators acting in nucleon Fock space. Some key relevant formulae are gathered for convenience in Appendix \ref{A}. 

In general the argument, $\psi(\ve{r})$, of creation and  destruction operators refers to the position of a nucleon relative to the origin of coordinates. It is therefore convenient to first change the integration in eq.(\ref{optwfdef}) to one over 3$A$ independent variables by introducing an extra integration over  $\ve{R}_A$ in the form
\beq \xi_{E^+}(\ve{r})=\sqrt{(A+1)}&&\!\!\!\!\!\!\!\int d\ve{\chi}_{1}\, \dots d\ve{\chi}_{(A-1)}d\ve{R}_A\delta(\ve{R}_A) \Psi^*_0(\ve{\chi}_1,\dots\,\ve{\chi}_{(A-1)})\eol&&\times \Psi_{E^+}(\ve{\chi}_1,\dots\,\ve{\chi}_{(A-1)},\ve{\chi}_{A}=\ve{r})\eol=\sqrt{(A+1)}&&\!\!\!\!\!\!\!\int d\ve{r}_{1}\, \dots d\ve{r}_{(A-1)}d\ve{r}_{A}\delta(\ve{R}_A) \Psi^*_0(\ve{r}_1,\dots\,\ve{r}_{A})\eol&&\times \Psi_{E^+}(\ve{r}_1,\dots\,\ve{r}_{A},\ve{r}_{(A+1)}=\ve{r}).\eol && \label{optwfdef2}\eeq
In the second equality  the variables of integration have been changed to $\ve{r}_1,\dots\,\ve{r}_{A}$ using the relation between these and the Jacobi coordinates for nucleons of the same mass:
\beq \ve{\chi}_{i}=\ve{r}_{(i+1)}-\ve{R}_i,\,\,\,i=1,\dots, A. \label{Jacobidef}\eeq
The inverse of these relations gives the set  $\ve{r}_1,\dots\,\ve{r}_{A}$ as linear functions of the set $\ve{\chi}_1,\dots\,\ve{\chi}_{(A-1)},\ve{R}_A.$ The coordinate $\ve{r}_{(A+1)}$ is given by $\ve{\chi}_A+\ve{R}_A=\ve{r}$ under the integral sign. (In understanding these  steps it may be helpful to consider the $A=2$ case explicitly, with $\ve{\chi}_1= (\ve{r}_2-\ve{r}_1),\ve{\chi}_2= (\ve{r}_3-\ve{R}_2),\, \ve{R}_2=(\ve{r}_2+\ve{r}_1)/2,$ and the inverse relations $ \ve{r}_3=\ve{R}_3+2\ve{\chi}_2/3, \ve{r}_2=\ve{R}_3-\ve{\chi}_2/3 +\ve{\chi}_1/2,\ve{r}_1=\ve{R}_3-\ve{\chi}_2/3 -\ve{\chi}_1/2.$ )

Note that strict notational conventions require the use of a different symbol for, \emph{e.g.}, $\Psi_0$, considered as a function of $\ve{r}_1,\dots\,\ve{r}_{A}$ in the last line in eq.(\ref{optwfdef2}). This step is avoided here by using the convention
  \beq \Psi_0 (\ve{r}_1,\dots\,\ve{r}_{A})=\Psi_0 (\ve{\chi}_1(\ve{r}_1,\dots\,\ve{r}_{A}),\dots\,\ve{\chi}_{(A-1)}(\ve{r}_1,\dots\,\ve{r}_{A})), \label{convention} \eeq

where $\ve{\chi}_i(\ve{r}_1,\dots\,\ve{r}_{A}),\, i=1,\dots, (A-1)$ refers to the linear equations of the transformation between the two sets of variables. 

The relation given in eq.(\ref{phiS}) of Appendix \ref{A} enables  the definition (\ref{optwfdef2}) to be written
 \beq \xi_{E^+}(\ve{r})=\la\la \Psi(0, \ve{x}=0) \mid \psi(\ve{r})\mid \Psi_{E^+}\ra\ra. \label{xiE+def}\eeq
where, for arbitrary $\ve{x}$,
\beq 
 \mid \Psi(n, \ve{x}) \rangle \rangle =\frac{1}{\sqrt{A!}}\int d\ve{r}_1\, d\ve{r}_2 \dots d\ve{r}_A \delta(\ve{R}_A-\ve{x})\Psi_n(\ve{r}_1, \dots, \ve{r}_A)\psi^\dagger(\ve{r}_A)\dots \psi^\dagger(\ve{r}_1)\mid 0 \rangle \rangle.\eol&&\label{Psi0xF}\eeq
 and
\beq 
 \mid \Psi_{E^+} \rangle \rangle &&\!\!\!\!\!\!=\frac{1}{\sqrt{(A+1)!}}\int d\ve{r}_1\, d\ve{r}_2 \dots d\ve{r}_{(A+1)} \Psi_{E^+}(\ve{r}_1, \dots, \ve{r}_{(A+1)})\psi^\dagger(\ve{r}_{(A+1)})\dots \psi^\dagger(\ve{r}_1)\mid 0 \rangle \rangle.\eol&&\label{PsiE+F}\eeq
 The kets $\mid \Psi(n, \ve{x}) \rangle \rangle$ form a complete set of antisymmetric $A$-nucleon states with  an intrinsic state labelled by $n$ and c.m. located at position $\ve{x}$ relative to a arbitrary origin. The bra in
  eq.(\ref{xiE+def}) has the c.m. of the $A$ target nucleons located at the origin, although in fact any value of $\ve{x}$ could have been chosen. 
  
  In this Section  a general definition of an optical model wave function, eq.(\ref{xiE+def}), has been obtained as the matrix element of a nucleon destruction operator between many-nucleon  states in Fock space by starting from a standard translationally invariant definition in terms of many-body wave functions expressed interns of Jacobi coordinates. In the next Section it is shown how the definition (\ref{xiE+def}) can be derived from more basic physical ideas of what is meant by the optical model wave function.     
 
\subsection{Physical basis for the optical model wave function.}\label{physicalbasis}
   The precise definition of the  $A+1$-nucleon scattering state $ \mid \Psi_{E^+}\rangle \rangle$ will be discussed  in Section \ref{Greenfunctions}. Leaving this aside for the present, a natural definition of the optical model wave function is the amplitude for finding in  $ \mid \Psi_{E^+}\rangle \rangle$, $A$ nucleons in the ground  state, $\psi_0(\ve{r}_1,\dots,\ve{r}_A)$, of the $A$-nucleon Hamiltonian with a  total momentum $\ve{K}'$, and a single nucleon at a distance $\ve{r}$ from the c.m. of the other $A$ nucleons.  This antisymmetric $A+1$-nucleon state is written $ \Phi_{\ve{r},0,\ve{K}'}$.  Note that this definition means that $\ve{r}$ is not the eigenvalue of a single nucleon position operator because the positions of all $A+1$ nucleons are involved in its definition.
    
     In configuration space  the state $ \phi_{\ve{r},0,\ve{K}'}$ is described by the wave function
\beq  \phi_{\ve{r},\,0,\ve{K}'}(\ve{r}_1,\dots,\ve{r}_A,\ve{r}_{(A+1)})\!\!\!\!\!\!&&=\frac{1}{\sqrt{(A+1)}}(1-\sum^A_{j=1}P_{((A+1),j)})\delta(\ve{r}_{(A+1)}-\ve{R}_A-\ve{r})\eol &&\times \frac{1}{(2\pi)^{3/2}}\exp(\imath \ve{R}_A.\ve{K}')\psi_{0}(\ve{r}_1,\dots,\ve{r}_A),\label{PsioK'r}\eeq  
where $P_{(i,j)}$ interchanges the all the coordinates of nucleons $i$ and $j$, and
\beq\ve{R}_A=(\sum_{i=1}^{i=A}\ve{r}_i)/A.\label{RAdef}\eeq
It is assumed that the many-nucleon Hamiltonian is translationally invariant and that the A-nucleon ground state $\psi_0$ has total momentum zero and is antisymmetrised in all nucleon coordinates. The factor $1/\sqrt{(A+1)}$ has been chosen for later convenience.

 For the wave function given in (\ref{PsioK'r}) the relation (\ref{SFock}) of Appendix \ref{A} gives the Fock space equivalent
 \beq 
 \mid \Phi(\ve{r},\,0,\ve{K}')\ra\ra \!\!\!\!\!\!&&=\frac{1}{\sqrt{A!}}\int d\ve{r}_1\, \dots d\ve{r}_A \frac{1}{(2\pi)^{3/2}}\exp(\imath \ve{R}_A.\ve{K}')\psi_{0}(\ve{r}_1,\dots,\ve{r}_A)\eol &&\times\psi^\dagger(\ve{r}+\ve{R}_A)\psi^\dagger(\ve{r}_A)\dots \psi^\dagger(\ve{r}_1)\mid 0 \rangle \rangle.\label{PsioK'r2}\eeq
 This state can be written in various ways. 
 
 Using
 \beq \exp(+\imath \ve{R}_A.\hat{\ve{P}})\psi^\dagger(\ve{r}_i) \exp(-\imath \ve{R}_A.\hat{\ve{P}})=\psi^\dagger (\ve{r}_i-\ve{R}_A),\label{PpsiP1}\eeq
and $\hat{\ve{P}}\mid 0 \ra\ra=0$,  eq.(\ref{PsioK'r2}) can be written
 \beq 
 \mid \Phi(\ve{r},\,0,\ve{K}')\ra\ra \!\!\!\!\!&&=\frac{1}{\sqrt{A!}}\int d\ve{r}_1\, \dots d\ve{r}_A\psi_{0}(\ve{r}_1,\dots,\ve{r}_A)\frac{1}{(2\pi)^{3/2}}\exp(\imath \ve{R}_A.(\ve{K}'-\hat{\ve{P}}))\eol
 &&\times\psi^\dagger(\ve{r})\psi^\dagger(\ve{r}_A-\ve{R}_A)\dots \psi^\dagger(\ve{r}_1-\ve{R}_A)\mid 0 \rangle \rangle.\label{PsioK'r3}\eeq 
 To exploit the fact that the ground state $\psi_{0}(\ve{r}_1, \dots \ve{r}_A)$ has zero momentum  the integration variables in eq.(\ref{PsioK'r3}) are changed to the set of $A-1$ translationally invariant variables $\ve{\chi}_1,\dots\,\ve{\chi}_{(A-1)}$ introduced following eq.(\ref{optwfdef}) above.  Together with $\ve{R}_A$ they form a set of variables equivalent to $\ve{r}_1, \dots \ve{r}_A$ and have a transformation Jacobian of $+1$. Under a translation of the coordinate system by $\ve{x}$ the $\ve{\chi}_1,\dots\,\ve{\chi}_{(A-1)}$ are unchanged  and $\ve{R}_A\rightarrow \ve{R}_A+\ve{x}$. 
 
The integral in eq.(\ref{PsioK'r3}) becomes
  \beq 
\mid \Phi(\ve{r},\,0,\ve{K}')\ra\ra \!\!\!\!\!&&=\frac{1}{\sqrt{A!}}\int d\ve{\chi}_{1}\, \dots d\ve{\chi}_{(A-1)} d \ve{R}_A\psi_0(\ve{r}_1,\dots,\ve{r}_A)\eol
 &&\times\frac{1}{(2\pi)^{3/2}}\exp(\imath \ve{R}_A.(\ve{K}'-\hat{\ve{P}}))\psi^\dagger(\ve{r})\psi^\dagger(\ve{r}_A-\ve{R}_A)\dots \psi^\dagger(\ve{r}_1-\ve{R}_A)\mid 0 \rangle \rangle.\eol &&\label{PsioK'r4}\eeq
 Note that in eq.(\ref{PsioK'r4}) the functional convention introduced in eq.(\ref{convention}) is used.

 The combinations $\ve{r}_i-\ve{R}$ that appear as arguments of the creation operators in eq.(\ref{PsioK'r4}) are all translationally invariant, as is $\psi_0(\ve{r}_1,\dots,\ve{r}_A)$. Therefore, they can be expressed entirely in terms of the variables $\ve{\chi}_1,\dots\,\ve{\chi}_{(A-1)}$ and are independent of $\ve{R}_A$. The integral over $\ve{R}_A$ can therefore be carried out to give
  \beq 
\mid \Phi(\ve{r},\,0,\ve{K}')\ra\ra \!\!\!\!\!\!\!\!\!&&=(2\pi)^{3/2}\delta(\ve{K}'-\hat{\ve{P}})\psi^\dagger(\ve{r})\eol&&\times[\frac{1}{\sqrt{A!}}\int d\ve{\chi}_{1}\, \dots d\ve{\chi}_{(A-1)} \psi_0(\ve{r}_1,\dots,\ve{r}_A)\psi^\dagger(\ve{r}_A-\ve{R}_A)\dots \psi^\dagger(\ve{r}_1-\ve{R}_A)\mid 0 \rangle \rangle],\eol &&\label{PsioK'r5}\eeq
 where the translationally invariant combinations $\ve{r}_i-\ve{R}_A,\,i=1\dots\, A$, and the arguments of $\psi_0$ are  functions  of the $(A-1)$ coordinates $\ve{\chi}_i$ only.
 
 The quantity in square brackets in eq.(\ref{PsioK'r5}) can be rewritten in terms of an integral over the $A$ coordinates 
 $\ve{r}_1, \dots \ve{r}_A$ by introducing an extra integration over $\ve{R}_A$ with a factor $\delta(\ve{R}_A) $ in the integrand as in the discussion around eq.(\ref{Psi0xF}). The quantity  in square brackets in eq.(\ref{PsioK'r5})  can now be written
 \beq \mid \Psi(0,\ve{x}=0)\ra\ra=&&\!\!\!\!\!\frac{1}{\sqrt{A!}}\int d\ve{\chi}_{1}\, \dots d\ve{\chi}_{(A-1)} \psi_0(\ve{r}_1,\dots\,\ve{r}_{A})\psi^\dagger(\ve{r}_A(\chi))\dots \psi^\dagger(\ve{r}_1(\chi))\mid 0 \rangle \rangle\eol &&\!\!\!\!\!\!\!\!\!\!\!\!\!\!\!=\frac{1}{\sqrt{A!}}\int d\ve{r}_{1}\, \dots d\ve{r}_{A}\delta(\ve{R}_A) \psi_0(\ve{r}_1,\dots\,\ve{r}_{A})\psi^\dagger(\ve{r}_A)\dots \psi^\dagger(\ve{r}_1)\mid 0 \rangle \rangle,\eol &&\label{psi00}\eeq
 In the first line $\ve{r}_i(\chi)$ means $\ve{r}_i-\ve{R}_A$ expressed as a function of the $(A-1)$ coordinates $\ve{\chi}_i$. 

 Comparing with eq.(\ref{SFock}) it can be seen that this a just the formula for the Fock space equivalent to the $A$-nucleon wave function $\delta(\ve{R}_A) \psi_0(\ve{r}_1,\dots\,\ve{r}_{A})$,  with an internal state characterised by the index "0" and zero momentum and with its c.m. located with certainty at the origin of coordinates.
 
The final expression for the state  $\mid \Phi(\ve{r},\,0,\ve{K}')\ra\ra$ needed to define the optical model wave function is therefore
  \beq 
\mid \Phi(\ve{r},\,0,\ve{K}')\ra\ra \!\!\!\!\!\!\!\!\!&&=(2\pi)^{3/2}\delta(\ve{K}'-\hat{\ve{P}})\psi^\dagger(\ve{r})\mid \Psi(0, \ve{x}=0)\ra\ra.\eol &&\label{PsioK'r6}\eeq

Using these definitions  the result for the overlap of this state with an $(A+1)$ nucleon state $\mid \Psi_{1,\ve{K}}\ra\ra$ with momentum $\ve{K}$ is
\beq \la\la\Phi(\ve{r},\,0,\ve{K}')\mid \Psi_{1,\ve{K}}\ra\ra\!\!\!\!\!\!\!\!\!&&=(2\pi)^{3/2}\delta(\ve{K}'-\ve{K})\la\la\Psi(0,\ve{x}=0)\mid\psi(\ve{r})\mid \Psi_{1,\ve{K}}\ra\ra,\eol &&\label{overlap}\eeq
where $\mid \Psi(0,\ve{x}=0)\ra\ra$ is defined in eq.(\ref{psi00}).

 The expression $\la\la \Psi(0, \ve{x}=0) \mid \psi(\ve{r})\mid \Psi_{1,\ve{K}}\ra\ra$ that appears on the right-hand-side of eq.(\ref{overlap}) is a more general example  of the overlap introduced earlier  in eq.(\ref{xiE+def}). Note that $\mid \Psi(0), \ve{x}=0)\ra\ra$ does not have definite momentum. In fact the $\delta(\ve{R}_A)$ factor in eq.(\ref{psi00}) means that in this state all values of the total momentum of the $A$-nucleons are equally probable. On the other hand  the operator $\delta(\ve{K}'-\hat{\ve{P}})$ in the ket $\mid \Phi(\ve{r},\,0,\ve{K}')$ given in eq.(\ref{PsioK'r5}) projects out a component of momentum $\ve{K}'$ from any state it acts on and hence gives rise to the momentum conserving delta function $\delta(\ve{K}'-\hat{\ve{K}})$ in the complete overlap $\la\la  \Phi_{\ve{r},\,0,\ve{K}'} \mid \Psi_{1,\ve{K}}\ra\ra$ given in eq.(\ref{overlap}). This delta function would be integrated over a narrow wave packet in momentum space in any complete scattering  theory. The other factors on the right-hand-side of eq.(\ref{overlap}) are the main focus of interest.
 
For translationally invariant Hamiltonians it can be assumed that  $\mid\Psi_{1,\ve{K}}\ra$  has the form
\beq \la \ve{r}_1,\dots,\ve{r}_{(A+1)}\mid\Psi_{1,\ve{K}}\ra=\frac{1}{(2\pi)^{3/2}}\exp(\imath \ve{R}_{(A+1)}.\ve{K})\Psi_1(\ve{r}_1,\dots,\ve{r}_{(A+1)}),\label{PsiK1}\eeq
where $\Psi_1$ is an intrinsic state of zero momentum.

 Scattering theories are usually expressed in terms of overall c.m. system quantities. In the following therefore  $\ve{K}=\ve{K}'=0$ is assumed  and any reference to these quantities is omitted  in the notation. The quantity  defined by
  \beq \xi_{0,1}(\ve{r})=\la\la \Psi(0, \ve{x}=0) \mid \psi(\ve{r})\mid \Psi_1\ra\ra. \label{xi01def}\eeq
  coincides with that introduced in Section \ref{FockSpaceformulae}, eq.(\ref{xiE+def}).
  
  It can be checked that if the overlap $\la\Phi(\ve{r},\,0,\ve{K}')\mid \Psi_{1,\ve{K}}\ra$ is calculated directly using the configuration space wave functions  (\ref{PsioK'r}) and (\ref{PsiK1}) one obtains the result corresponding to (\ref{optwfdef}) after a momentum conserving delta function is removed and $\ve{K}$ and $\ve{K}'$ are set equal to zero.
  
  In the next Section it will be shown that when $\Psi_0$ and $\Psi_1$ are associated with the same Hamiltonian operator in Fock space, $\xi_{0,1}$ satisfies an inhomogeneous differential equation ("source equation") with a kinetic energy term that carries the correct reduced mass for motion of one nucleon relative to the c.m. of the other $A$ nucleons.
  \section{Source equation for $\xi_{0,1}(\ve{r})$.}\label{sourceequation}
It is assumed that $\psi_0(\ve{r}_1,\dots,\ve{r}_A)$ is an eigenstate of the $A$-nucleon intrinsic Hamiltonian $H-(\ve{P})^2/(2Am)$ with eigenvalue $E_0$, where $\ve{P}$ is the $A$-nucleon total momentum operator. By construction the intrinsic Hamiltonian operator commutes with the c.m. coordinate $\ve{R}_A$ and hence 
\beq (\hat{H}-(\ve{\hat{P}})^2/(2Am))\mid \Psi(0, \ve{x}=0) \ra\ra=E_0\mid \Psi(0, \ve{x}=0),
 \ra\ra.\label{HPsi00}\eeq
 is a valid Fock-space equation where $\hat{H}$ and  $\ve{\hat{P}}$ are Fock-space operators and $\mid \Psi(0, \ve{x}=0) \ra\ra$ is the state defined in eq.(\ref{psi00}).
  
If $\Psi_1$ is an $(A+1)$, zero momentum, eigenstate  of $H$ with eigenvalue $E_1$ it is also an eigenstate of the Fock-space operator $ (\hat{H}-(\ve{\hat{P}})^2/(2Am))$ because $\ve{\hat{P}}\mid \Psi_1\ra\ra=0$.
\beq (\hat{H}-(\ve{\hat{P}})^2/(2Am))\mid \Psi_1 \ra\ra=E_1\mid \Psi_1 \ra\ra.\label{HPsi1}\eeq
It follows that
\beq(E_0-E_1)\xi_{0,1}(\ve{r})&&\!\!\!\!\!\!\!=\la\la \Psi(0,\ve{x}=0) \mid (\hat{H}-(\ve{\hat{P}})^2/(2Am)))\psi(\ve{r})\mid \Psi_1\ra\ra\eol &&-\la\la \Psi(0,\ve{x}=0) \mid\psi(\ve{r})((\hat{H}-(\ve{\hat{P}})^2/(2Am))\mid \Psi_1\ra\ra\eol&&\!\!\!\!\!\!\!\!\!=\la\la \Psi(0,\ve{x}=0) \mid[(\hat{H}-(\ve{\hat{P}})^2/(2Am)),\psi(\ve{r})]_-\mid \Psi_1\ra\ra. \label{sourcederivation1}\eeq
It is  assumed that $\hat{H}$ can be written as the sum of a kinetic energy term $\hat{T}$ and a potential energy term $\hat{V}$:
\beq \hat{H}=\hat{T} + \hat{V}, \label{H} \eeq
 where
 \beq \hat{T}= -\frac{\hbar^2}{2m}\int d\ve{r}'\psi^\dagger(\ve{r}')(\nabla^2_{\ve{r}'}\psi(\ve{r}')),\label{T}\eeq
 and hence
 \beq [\hat{T},\psi^\dagger(\ve{r})]_- &=& -\frac{\hbar^2}{2m}(\nabla^2_{\ve{r}}\psi^\dagger(\ve{r})), \label{Tcom1}\eeq
and
\beq [\hat{T},\psi(\ve{r})]_- &=&+\frac{\hbar^2}{2m}(\nabla^2_{\ve{r}}\psi(\ve{r})). \label{Tcom2}\eeq

Using these commutators and the results (\ref{Parcomm}) in eq.(\ref{sourcederivation1})  gives
\beq(E_0-E_1)\xi_{0,1}(\ve{r})&&\!\!\!\!\!\!\!\!\!=\la\la \Psi(0,\ve{x}=0) \mid[\hat{T},\psi(\ve{r})]_-\mid \Psi_1\ra\ra+\la\la \Psi(0,\ve{x}=0) \mid[\hat{V},\psi(\ve{r})]_-\mid \Psi_1\ra\ra\eol &&-\la\la \Psi(0,\ve{x}=0) \mid[(\ve{\hat{P}})^2/(2Am),\psi(\ve{r})]_-\mid \Psi_1\ra\ra\eol &&\!\!\!\!\!\!\!\!\!=\la\la \Psi(0,\ve{x}=0) \mid\frac{\hbar^2}{2m}(\nabla^2_{\ve{r}})\psi(\ve{r})\mid \Psi_1\ra\ra+\la\la \Psi(0,\ve{x}=0) \mid[\hat{V}, \psi(\ve{r})]_-\mid \Psi_1\ra\ra\eol &&-\la\la \Psi(0,\ve{x}=0) \mid\frac{-\hbar^2}{2mA}(\nabla^2_{\ve{r}})\psi(\ve{r})\mid \Psi_1\ra\ra\eol &&\!\!\!\!\!\!\!\!\!=(1+\frac{1}{A})(\frac{\hbar^2}{2m}\nabla^2_{\ve{r}})\xi_{0,1}(\ve{r})+\la\la \Psi(0,\ve{x}=0) \mid[\hat{V}, \psi(\ve{r})]_-\mid \Psi_1\ra\ra. \eol &&\label{sourcederivation2}\eeq
Hence 
\beq(-\frac{\hbar^2}{2\mu_{mA}}\nabla^2_{\ve{r}}-(E_1-E_0))\xi_{0,1}(\ve{r})&=&\la\la \Psi(0,\ve{x}=0) \mid[ \psi(\ve{r}),V]_-\mid \Psi_1\ra\ra, \eol &&\label{sourcederivation3}\eeq
where
\beq \mu_{mA}=\frac{A}{(A+1)}m, \label{redm}\eeq
 is the nucleon-$A$ reduced mass.

In the case that  $\Psi_1$ describes a scattering state in the overall c.m. system corresponding to elastic scattering of a nucleon by an $A$-nucleon in its ground state $\Psi_0$, the eigenvalue  $E_1$ is
\beq E_1=E_0+\frac{\hbar^2k^2_1}{2\mu_{mA}},\label{E1}\eeq
where $\ve{k}_1$ is the incident nucleon momentum in the c.m. system.

Eq.(\ref{sourcederivation3}) can now be written
\beq(-\frac{\hbar^2}{2\mu_{mA}}\nabla^2_{\ve{r}}-\frac{\hbar^2k^2_1}{2\mu_{mA}})\xi_{0,1}(\ve{r})&=&\la\la \Psi(0,\ve{x}=0) \mid[ \psi(\ve{r}),V]_-\mid \Psi_1\ra\ra, \eol &&\label{sourcederivation4}\eeq

Many techniques have been developed to deal with the  source term  $[\psi(\ve{r}),V]_-$, including work dedicated to expressing the source term as an operator in single nucleon degrees of freedom acting on $\xi_{0,1}(\ve{r})$ and hence providing a microscopic basis for the optical model. The aim of much of this work is to link up with standard many-body theories of nuclear structure which exploit a basis of determinants of single nucleon states in a potential fixed relative to the origin of coordinates and hence introduce a violation of translation invariance very early in the development. For example, this is true of the work of Bell and Squires \cite{BellSquires} and Capuzzi and Mahaux\cite{CapuzziMahaux95}, as well as  the more recent work\cite{Rotureau16} cited in the Introduction.  In common with references  \cite{BellSquires}, Capuzzi and \cite{CapuzziMahaux95}, and \cite{Rotureau16} they use a definition of the optical model wave function based eq.(\ref{chirdef}) instead of eq.(\ref{xi01def}). This is at least formally unacceptable. Further work is needed to assess any quantitative consequences.

As a first step to carrying forward an approach based on eq.(\ref{sourcederivation4}),  the next Section shows how the modified definition of the optical model wave function given in eqs.(\ref{xiE+def}) and (\ref{xi01def}) can be expressed in terms of a time dependent one-body Green's function. It was mentioned in the Introduction that time dependent Green's function methods are believed to be the way forward for medium and heavy targets. 
\section{The optical model wave function in terms of a time dependent Green's function.}\label{Greenfunctions}
 $\Psi_1$ describes a scattering state with a plane wave nucleon incident  on an $A$-nucleon ground state $\psi_0$ as the incident channel and outgoing waves in all other channels. In the overall c.m. system this state is the limit as $\epsilon \rightarrow 0$ of the state
\beq\mid \Psi^\epsilon_1\ra\ra =\frac{\imath \epsilon}{E_1-H+\imath \epsilon}(2\pi)^{3/2}a^\dagger_{\ve{k}_1}\mid -\ve{k}_1,\psi_0 \ra\ra, \label{Psi1scatt}\eeq
where $(2\pi)^{3/2}a^\dagger_{\ve{k}_1}\mid -\ve{k}_1,\psi_0 \ra\ra$ describes the incident channel  in which  the incident nucleon has momentum $\ve{k}_1$  and the  target has a total momentum $-\ve{k}_1$. In configuration space this incident channel state is the antisymmetrised version of
\beq \exp(\imath \ve{k}_1.\ve{\chi}_A)\psi_0(\ve{\chi}_1,\dots,\ve{\chi}_{(A-1)})
=\exp(\imath \ve{k}_1.\ve{r}_{(A+1)})\exp(-\imath \ve{k}_1.\ve{R}_{A})\psi_0(\ve{r}_1,\dots,\ve{r}_{A}).\label{incidentPW}\eeq
The expression (\ref{Psi1scatt}) is the generalisation to include c.m. degrees of freedom explicitly of the formulation of collision theory described in \cite{CapuzziMahaux95}, their eq.(4.19), and\cite{Villars1967}.

In order to make a connection with time-dependent Green functions  a time variable is introduced via the identity ($\hbar=1$)
\beq  \frac{\imath \epsilon}{(E_1+\imath \epsilon-H)}=\epsilon \int_{-\infty}^0dt\,\exp (-\imath(E_1+\imath \epsilon -H)t).
\label{identity}\eeq
The scattering state defined in eq.(\ref{Psi1scatt}) can now be written
\beq \mid \Psi^\epsilon_1\ra\ra&=&\epsilon \int_{-\infty}^0dt\,\exp (-\imath(E_1+\imath \epsilon -H)t)\eol
&\times &(2\pi)^{3/2}a^\dagger_{\ve{k}_1}\mid -\ve{k}_1,\psi_0 \ra\ra,\label{Psi1scatt1}\eeq
and  from eq.(\ref{xi01def}) the optical model overlap is
\beq \xi^\epsilon_{0,1}(\ve{r})&&=\la\la \Psi(0,\ve{x}=0) \mid\psi(\ve{r})\epsilon \int_{-\infty}^0dt\,\exp (-\imath(E_1+\imath \epsilon -H)t)\eol
&&\times (2\pi)^{3/2}a^\dagger_{\ve{k}_1}\mid -\ve{k}_1,\psi_0 \ra\ra. \label{xi01def2}\eeq

This expression can be written in terms of  the Heisenberg operators
\beq \Psi(\ve{r},t)&=&\exp(\imath H\,t)\psi(\ve{r})\exp(-\imath H\,t)
\eol
\Psi^{\dagger}(\ve{r},t)&=&\exp(\imath H\,t)\psi^{\dagger}(\ve{r})\exp(-\imath H\,t)\eol
A^{\dagger}_{\ve{k}_1}(t)&=&\exp(\imath H\,t)a^{\dagger}_{\ve{k}_1}\exp(-\imath H\,t).\label{Heisenberg} \eeq
 
In terms of these eq.(\ref{xi01def2} becomes
\beq \xi^\epsilon_{0,1}(\ve{r})&&=\la\la \Psi(0, \ve{x}=0) \mid\epsilon \int_{-\infty}^0dt\,\exp (-\imath(E_1+\imath \epsilon)t)\eol
&\times &(2\pi)^{3/2}\exp (\imath Ht)a^\dagger_{\ve{k}_1}\exp (-\imath Ht)\times \exp (+\imath Ht)\mid -\ve{k}_1,\psi_0 \ra\ra\eol &&=\la\la \Psi(0, \ve{x}=0) \mid \epsilon \int_{-\infty}^0dt\,\Psi(\ve{r},t=0)\exp (-\imath(E_1+\imath \epsilon )t)\eol
&&\times (2\pi)^{3/2}A^\dagger_{\ve{k}_1}(t)\exp(\imath( E_0+E_{1A})\,t)\mid -\ve{k}_1,\psi_0 \ra\ra\eol &&
=\la\la \Psi(0, \ve{x}=0) \mid \epsilon \int_{-\infty}^0dt\,\Psi(\ve{r},t=0)\exp (-\imath(\epsilon_{\ve{k_1}}+\imath \epsilon )t)\eol
&&\times (2\pi)^{3/2}A^\dagger_{\ve{k}_1}(t)\mid -\ve{k}_1,\psi_0 \ra\ra. \label{xi01def4}\eeq
The derivation of the last line of eq.(\ref{xi01def4}) has used
\beq H\mid -\ve{k}_1,\psi_0 \ra\ra=(E_{0}+E_{1A})\mid -\ve{k}_1,\psi_0 \ra\ra \label{EoE1Adef}\eeq
where $E_{0}$ is the ground state energy of the target and $E_{1A}$ is the incident target c.m. kinetic energy
\beq E_{1A}=\frac{\hbar^2k^2_1}{2Am}. \label{E1A} \eeq
 $E_1- E_{0}-E_{1A}$ is therefore the  incident nucleon kinetic energy 
\beq\epsilon_{\ve{k_1}}=\frac{\hbar^2k^2_1}{2m}.\label{epsilonk1}\eeq 

The expression (\ref{xi01def4}) for the optical model wavefunction can be written in terms of a Green function by using
\beq a^{\dagger}_{\ve{k}}=\int\,d\ve{r}'\frac{\exp(\imath \ve{k}.\ve{r}')}{(2\pi)^{3/2}}\psi^{\dagger}(\ve{r}').\label{kr} \eeq
It follows from this that
\beq a^{\dagger}_{\ve{k}_1}&&=\int d\ve{k}'\delta(\ve{k}_1-\ve{k}')a^{\dagger}_{\ve{k}'}\eol 
&&=\int d\ve{k}'(\int\,d\ve{r}'\frac{\exp(\imath (\ve{k}_1-\ve{k}').\ve{r}')}{(2\pi)^{3}})a^{\dagger}_{\ve{k}'}\eol&&=
\int\,d\ve{r}'\frac{\exp(\imath \ve{k}_1.\ve{r}')}{(2\pi)^{3/2}}\int d\ve{k}'\frac{\exp(-\imath \ve{k}').\ve{r}'}{(2\pi)^{3/2}}a^{\dagger}_{\ve{k}'}\eol&&=
\int\,d\ve{r}'\frac{\exp(\imath \ve{k}_1.\ve{r}')}{(2\pi)^{3/2}}\psi^{\dagger}(\ve{r}').\label{apsi} \eeq
and hence
\beq A^\dagger_{\ve{k}_1}(t)&&=\exp(\imath H\,t)a^{\dagger}_{\ve{k}_1}\exp(-\imath H\,t)\eol&&=
 \int\,d\ve{r}'\frac{\exp(\imath \ve{k}_1.\ve{r}')}{(2\pi)^{3/2}}\Psi^{\dagger}(\ve{r}',t).\label{AtPsit}\eeq
It follows from eqs.(\ref{xi01def4}) and (\ref{AtPsit}) that
\beq \xi^\epsilon_{0,1}(\ve{r})&&
=\la\la \Psi(0, \ve{x}=0) \mid\epsilon \int_{-\infty}^0dt\,\Psi(\ve{r},t=0)\exp (-\imath(\epsilon_{\ve{k_1}}+\imath \epsilon )t)\eol
&&\times (2\pi)^{3/2}A^\dagger_{\ve{k}_1}(t)\mid -\ve{k}_1,\psi_0 \ra\ra,\eol 
&&=\epsilon \int_{-\infty}^0dt\int\,d\ve{r}'\,\exp(-\imath(\epsilon_{\ve{k_1}}+\imath \epsilon)\,t))\eol &&\times \exp(\imath \ve{k}_1.\ve{r}')\langle   \Psi(0, \ve{x}=0)\mid \mathcal{T}\{\Psi(\ve{r},0)\,\Psi^{\dagger}(\ve{r}',t)\}\mid -\ve{k}_1,   \psi_0 \rangle.\eol &&
\label{xi01def5}\eeq
The time ordering operator $\mathcal{T}$ can be introduced without error because all values of $t$ in the integral satisfy $t<0$. 
 
Eq.(\ref{xi01def5}) can be written
\beq \xi^\epsilon_{0,\ve{k}_1}(\ve{r})&=&\epsilon \int_{-\infty}^0 dt \int\,d\ve{r}'\,\exp(\imath(\ve{k}_1.\ve{r}'-(\epsilon_{\ve{k_1}}+\imath \epsilon)\,t)) \eol &\times & G(\ve{r},t=0;\ve{r}',t)
\label{chiopt4}\eeq
where the Green function is given by
\beq  G(\ve{r},t;\ve{r}',t')=\langle \Psi(0, \ve{x}=0)\mid \,
\mathcal{T}\{\Psi(\ve{r},t)\,\Psi^{\dagger}(\ve{r}',t')\}\mid  -\ve{k}_1,   \psi_0\rangle.
\label{G}\eeq
This differs from the usual groundstate-groundstate one-nucleon Green function in that in the bra the groundstate has a c.m. localised at the origin and the ket the ground state has the total momentum $-\ve{k}_1$, \emph{i.e.,} opposite to the incident nucleon momentum in the overall c.m. system.
It would be interesting to explore how far this differences leads to significant quantitative effects on the calculation of microscopic optical potentials.

Eq.(\ref{chiopt4}) differs from the expression for the optical model wavefunction given by Bell and Squires \cite{BellSquires}. These differences appear to come from the way boundary conditions are handled. Bell and Squires introduce an unusual asymptotic condition which specifies a point source of particles at a large distance. They then project out a state with definite energy by integrating over all times. The scattering state (\ref{Psi1scatt}) is based on a physically transparent and well understood limiting process. The state $\mid \Psi^{(\epsilon)}_{\ve{k} }\rangle$  defined in Eq.(\ref{Psi1scatt}) is the state at time $t=0$ that evolves from a state in the remote past with energy $E$ with a spread of $\epsilon$ (see Gellman and Goldberger\cite{goldberger}). It is not clear at this time how these two different ways of handling the time effect the resulting theoretical optical potentials. They clearly involve two different time orderings in the Green function (\ref{G}) and this may alter the diagrams appearing in the perturbation expansion of the Green function.

Extensive quantitative comparisons of translational invariant and non-invariant calculations of overlap functions of bound state wave functions ("center-of-mass corrections") have been published in \cite{Tim11}, and references therein. These calculations evaluate the source term in eq.(\ref{sourcederivation4}) within the fully antisymmetrised translationally invariant oscillator shell-model and use a formulation of the overlap function in terms of Jacobi coordinates as defined in eq.(\ref{optwfdef}) of Section \ref{optwfdef1}. Therefore, they do not suffer from the definition difficulties set out in Section \ref{why}.  

Of particular relevance are the results obtained in \cite{Tim11} for Asymptotic Normalisation Coefficients (ANCs), which determine the amplitude of the overlap function in the region outside the $A$-nucleon nucleus and which one might expect are influenced by similar considerations as the optical model scattering amplitude in the case of an optical model overlap.
Ref. \cite{Tim11}, Table 1, quotes c.m. corrections to squared ANCs of $15.5\%$ even for the $^{40}$Ca-$^{41}$Ca overlap with $A$ as large as 40. According to ref. \cite{Tim11} these corrections are much larger than one would expect simply from the use of the correct reduced mass in the kinetic energy term in the source equation and scaling corrections deduced from the harmonic oscillator model. This suggests that similar quantitative effects may arise from c.m. corrections  incorporated in the new formalism developed in Sections \ref{sourceequation} and \ref{Greenfunctions} for the case of nucleon-nucleus scattering.
\section{Conclusions}
A new fully antisymmetrised, translationally invariant definition of the nucleon-nucleus optical model wave function in terms of many-nucleon scattering wave function has been introduced. It has been shown that this wavefunction satisfies a differential equation in which the kinematically correct kinetic energy operator appears. It has also been shown how this wave function can be related to a modified definition of a one-nucleon time-dependent Green's function.
\section{Acknowledgements.}
 This work has benefited from discussions with several members of the Surrey Nuclear Theory Group:  C.Barbieri, A.Idini, P.D. Stevenson and N.K. Timofeyuk. Support from the UK Science and Technology Facilities Council through the grant STFC ST/000051/1 is acknowledged.

\appendix
\section{Brief notes on 2nd Quantisation.}\label{A}
Many-body theory can be formulated in a new space (Fock space) in which states exist that have components in different orthogonal sub-spaces, each of which corresponds to a different number of particles (nucleons in the present case), including one component with zero particles, the vacuum. The Fock space concept allows the introduction of operators, such as creation and destruction operators that act in Fock space and connect sub-spaces with different numbers of particles. The nuclear states dealt with here usually have a definite number of particles and hence only one component in Fock space, and Hamiltonians that do not connect states with different numbers of particles, but in handling antisymmetry requirements Fock space ideas frequently simplify calculations considerably. In BCS pairing theory one actually does deal with states that do not have a definite number of nucleons and therefore have non-zero components in orthogonal parts of Fock space.

A general state vector, $\mid A \rangle\rangle$, in Fock space has the form
\beq
\mid A \rangle\rangle= \left \{ \begin{array}{c}
A_0  \\ \mid A_1\rangle_{(1)}\\ \mid A_2\rangle_{(1)(2)} \\ .\\.\\ \mid A_n\rangle_{(1)(2)\dots (n)} \\ .\\.  
 \end{array}\right \},\label{FockAB}\eeq
where the number $A_0$ is the amplitude for finding the system in a state with zero nucleons. The state with zero nucleons is called the vacuum and has a zero in every row except the first, \emph{i.e.}
 \beq
\mid 0 \rangle\rangle= \left \{ \begin{array}{c}
1 \\ 0\\ 0 \\ .\\.    
 \end{array}\right \}.\label{vacuum}\eeq
 In eq.(\ref{FockAB}) the quantity $\mid A_1\rangle_{(1)}$
is a state vector in the single nucleon subspace of nucleon labelled "1" , $\mid A_2\rangle_{(1)(2)}$ is an antisymmetrized state in the 2-nucleon subspace of nucleons "1" and "2", and, in the row labelled $n$, $\mid A_n\rangle_{(1)(2)\dots (n)}$ is an antisymmetrized state in the $n$-nucleon subspace of nucleons "1" to "$n$".

Note that  the rows of  $\mid A \rangle\rangle$ are labelled with the integers $0,1,2,\dots$, starting with row "0" which contains the vacuum amplitude $A_0$.

In general a double right-angle bracket, $\mid\dots \rangle \rangle$, is used to denote state vectors in Fock space and the usual single right-hand bracket, $\mid\dots \rangle $, to denote a state with a definite number of nucleons. 
\subsection{Definition of creation and destruction operators.}
The complete definition of the creation and destruction operators acting on a general ket $\mid A \rangle \rangle$ in Fock space as ( $x$ signifies a set of commuting nucleonic dynamic variables for space, spin and isospin)
\beq
\psi^\dagger(x_0)\mid A \rangle\rangle&=& \left \{ \begin{array}{c}
0  \\ A_0\mid x_0\rangle_{(1)}\\\frac{1}{\sqrt{2}} \mathcal{A}_{1(2)}[\mid x_0\rangle_{(1)}\mid A_1\rangle_{(2)} ]\\ .\\.\\ \frac{1}{\sqrt{n}}\mathcal{A}_{1(2\dots n)}[\mid x_0 \rangle_{(1)}\mid A_{n-1}\rangle_{(2)\dots (n)}]   \\ .\\.
 \end{array}\right \},\eol\psi(x_0)\mid A \rangle\rangle&=& \left \{ \begin{array}{c}
\langle x_0\mid A_1 \rangle  \\ \sqrt{2}\int dx_2\mid x_2 \rangle_{(1)}\langle x_0, x_2\mid A_2\rangle\\  .\\.\\ \sqrt{(n+1)}\int dx_2\dots dx_{n+1}\mid x_2,\dots x_{n+1}\rangle_{(1)\dots (n)} \langle x_0,x_2\dots x_{n+1}  \mid A_{n+1}\rangle \\ .\\.  
 \end{array}\right \}.\eol&&
 \label{psidaggerpsiA}\eeq
 Here $\mathcal{A}_{1(2\dots n)}$ acts on the labels of nucleons $1,2,\dots n,$ and is defined by
\beq \mathcal{A}_{1(2\dots n)}&=& 1-\sum_{j=2}^{j=n}P_{(1,j)},\eol &&\label{Anti}\eeq
where $P_{(1,j)}$ interchanges the labels of nucleons $1$ and $j$.
\subsection{Connection between a state in Fock space with a definite number of nucleons and a many -body wave function}
 An $A$-nucleon state with wave function $\phi_S(x_1, x_2 \dots x_A)$ is described in Fock space by a vector $ \mid \Phi_S \rangle \rangle $  given by
\beq 
 \mid \Phi_S \rangle \rangle =\frac{1}{\sqrt{A!}}\int dx_1\, dx_2 \dots dx_A \phi_S(x_1, \dots x_A)\psi^\dagger(x_A)\dots \psi^\dagger(x_1)\mid 0 \rangle \rangle.\label{SFock}\eeq
  The notation used is $x_1=\ve{r}_1, s_z,\tau_3$, \emph{etc.}., and the integral signs include a summation over $s_z=\pm 1/2, \tau_3=\pm 1/2.$

 The inverse of (\ref{SFock}) is
 \beq 
 \phi_S(x_1, \dots x_A) =\frac{1}{\sqrt{A!}}\langle \langle 0 \mid\psi(x_1)\dots \psi(x_A)\mid \Phi_S \rangle \rangle.\label{phiS}\eeq

 The operator $\psi^\dagger(x_0)$ creates a particle at $x_0$. Its adjoint  $\psi(x_0)$ destroys a particle at point $x_0$. A formal definition of what these statements mean in terms of the action of these operators on an arbitrary Fock state is given in eqs.(\ref{psidaggerpsiA}) above; however, these definitions are rarely needed in practice and all one invariably has to use is the anti-commutation relations that can be derived from the definitions:
\beq \psi^\dagger(x_0)\psi^\dagger(x_0')+\psi^\dagger(x_0')\psi^\dagger(x_0) &=&0,\,\,\,\psi^\dagger(x_0)\psi^\dagger(x_0)=0 \eol
\psi(x_0)\psi(x_0')+\psi(x_0')\psi(x_0) &=&0,\,\,\,\psi(x_0)\psi(x_0)=0\eol\psi(x_0)\psi^\dagger(x_0')+\psi^\dagger(x_0')\psi(x_0)&=&\delta(x_0-x_0').\label{psiprops}\eeq


\begin{thebibliography}{} 
   \bibitem{RQN14} C.Romero-Redondo, S.Quaglioni and P.Navratil, Phys.Rev.Letts.113, 032503(2014).
   \bibitem{NQ12} P.Navratil and S.Quaglioni, Phys.Rev.Letts.108, 042503(2012).
 \bibitem{IdiniBarbieriNavratil16}A.Idini, C.Barbieri and P.Navratil, arXiv:1612.01478v1 [nucl-th] 5 Dec 2017.
13 (2011).
\bibitem{Fes58} H.Feshbach, Ann. Phys. 5, 357 (1958)
\bibitem{BellSquires} J. Bell and E.Squires, Phys. Rev. Letts.3, 96(1959).
\bibitem{CapuzziMahaux95}F.Capuzzi and C.Mahaux, Annals of Physics (NY)239, 57(1995).
\bibitem{Rotureau16}J.Rotureau, P. Danielewicz, G.Hagen, F.M. Nunes and T.Papenbrock, Phys.Rev.C95, 024315(2017)
\bibitem{Tim11} N.K. Timofeyuk, Phys.Rev. C 84, 054313 (2011).
.\bibitem{Tim14} N.K. Timofeyuk, J.Phys.G:Nucl.Part.Phys.41,094008(2014).
\bibitem{Villars1967}F.Villars,  "Collision Theory", in \emph{Fundamentals in Nuclear Theory}, Lecture Notes Presented at an International Course, Trieste, 3 Oct.-16 Dec. 1966, organised by te International Centre for Theoretical Physics, Trieste, Ed. by A. de-Shalit and C.Villi, (Int.Atomic Energy Agency, Vienna 19670)ST1/PUB?145, pages 260-333.

\bibitem{goldberger} M.L. Goldberger and K.M. Watson, {\it Collision Theory}, 
Wiley, New York 1964 (Dover Publications, N.Y, 2004).
\end{thebibliography}
 \end{document}